\documentclass[aps,prb,twocolumn,showpacs,superscriptaddress,floatfix]{revtex4-1}

\usepackage{amsmath}
\usepackage{amssymb}
\usepackage{graphicx}
\usepackage{grffile} 
\usepackage{float}
\usepackage{color}
\usepackage{subfigure}
\usepackage[normalem]{ulem}

\newcommand{\kk}{\mathbf{k}}

\newcommand{\A}{\mathbf{A}}

\newcommand{\LK}[1]{\textcolor{black}{#1}}

\renewcommand{\sout}[1]{}

\begin{document}

\author{A.~F.~Kemper}
\email{akemper@ncsu.edu}
\affiliation{Department of Physics, North Carolina State University, Raleigh, NC 27695, USA}
\affiliation{Lawrence Berkeley National Laboratory, 1 Cyclotron Road, Berkeley, CA 94720, USA}

\author{M.~A.~Sentef}
\affiliation{HISKP, University of Bonn, Nussallee 14-16, D-53115 Bonn, Germany}

\author{B.~Moritz}
\affiliation{Stanford Institute for Materials and Energy Sciences (SIMES),
SLAC National Accelerator Laboratory, Menlo Park, CA 94025, USA}
\affiliation{Department of Physics and Astrophysics, University of North Dakota, Grand Forks, ND 58202, USA}

\author{J.~K.~Freericks}
\affiliation{Department of Physics, Georgetown University, Washington, DC 20057, USA}

\author{T.~P.~Devereaux}
\affiliation{Stanford Institute for Materials and Energy Sciences (SIMES),
SLAC National Accelerator Laboratory, Menlo Park, CA 94025, USA}
\affiliation{Geballe Laboratory for Advanced Materials, Stanford University, Stanford, CA 94305}

\pacs{ }

\title{Direct observation of Higgs mode oscillations in the pump-probe photoemission spectra of electron-phonon mediated superconductors}

\begin{abstract}
\LK{
Using the non-equilibrium Keldysh formalism, we solve the equations of motion for electron-phonon superconductivity, including an ultrafast pump field.
We present results for time-dependent photoemission spectra out of equilibrium which probes the dynamics of the superconducting gap edge.
The partial melting of the order by the pump field leads to oscillations at twice the melted gap frequency, a hallmark of the Higgs or amplitude mode. 
Thus the Higgs mode can be directly excited through the nonlinear effects of an electromagnetic field and detected without any additional symmetry breaking.
}
\end{abstract}

\maketitle

\section{Introduction}
The amplitude, or Higgs mode is deeply intertwined with the historical development of the BCS theory of superconductivity.
Although the presence of the Higgs mode is fundamental to superconductivity, it remained elusive for many
decades,
and its presence and observability is still under debate in many contexts.\cite{tsuchiya_13,pekker_14,cea_14}
Direct observation of the Higgs mode as an oscillation of the superconducting order parameter
is difficult with standard methods 
since it does not couple linearly to electromagnetic fields 
\cite{podolsky_11}.
It can be observed
if it is coexistent with another order, such as a charge
density wave\cite{sooryakumar_80, littlewood_81, measson_14}, however it is difficult to distinguish from
other effects such as phonon splitting due to the secondary order\cite{measson_14}.

More generally, the dynamics of superconductors is a field of study with a long rich history. 
Until recently,
studies were mainly limited to the frequency domain, where measurements
are averaged over long times. This changed with the advent of time-resolved spectroscopy, which
is performed by exciting the system with an ultrashort pump laser pulse, followed by an equally short
probe pulse. 
\sout{
The most common measurements of this type are optical spectroscopy, either in
transmission or in reflection, and more recently time- and angle-resolved photoemission
spectroscopy.  The second
is proving to be a particularly powerful tool due to its ability to resolve the single-particle spectral
function, which exhibits many characteristics of the underlying system, over time.
}  
These tools have
opened a new window into the complex dynamics of superconductors
(as well as other ordered phases, e.g. CDW insulators\cite{s_hellmann_12})
by performing studies on the 
gap\cite{beck_11,smallwood_14}, 
collective\cite{hinton_13},
quasiparticle\cite{cortes_11,smallwood_12}, 
and interaction dynamics\cite{rettig_13,rameau_14,zhang_14,dalconte_15}.

Theoretical studies of the Higgs mode in the time domain have a longer history than experiments
\sout{ 
Already early on,
theoretical studies of dynamics after a perturbation were performed, showing various types of
dynamic excitations} 
(see e.g. Refs.~\onlinecite{volkov_74,kulik_81,yuzbashyan_06}).  
However, the theory is usually
done within the context of BCS theory, and more importantly, focuses on single-time dynamics.
These neglect important dynamical processes that are present in real materials, including
melting of the superconducting order, and relaxation processes, and thus provide
a qualitative description at best.
Recently, it was shown
that a driving field tuned to a frequency $2 \omega=2\Delta$, twice the superconducting gap, could
be used to resonantly excite the amplitude mode in NbN\cite{tsuji_14,matsunaga_13,matsunaga_14},
which was confirmed by THz transmission experiments.

In this work, we will show that
the amplitude mode can be excited without tuning to a resonance,
but rather that it is a fundamental part of the \sout{post-pump} dynamics, 
and how time- and angle-resolved photoemission spectroscopy (tr-ARPES) 
can be used to directly observe
the amplitude mode, without the need for an additional degree of freedom for the amplitude mode
to couple to.
\sout{The process can be illustrated as follows (see Fig.~\ref{fig:mexicanhat}). }
The pump pulse perturbs the ordered state,
resulting in
changes of the effective free energy landscape $\mathcal{F}$.
The minimum in $\mathcal{F}$ is reduced and shifts towards smaller order $|\Delta|$.
\sout{due to the decrease of quasiparticles involved in ordering.}
This goes beyond the linear response regime, where the system is simply perturbed
from its equilibrium state without affecting the free energy landscape in which it lives.
The non-linear coupling is critical to the excitation of the Higgs mode.\sout{, as we will show below.}
\sout{If the response of the complex order parameter ($\Delta e^{i\phi}$) is 
slower than the changes in $\mathcal{F}$, }
After perturbation, oscillations about
the new minimum will occur at a frequency of $2|\Delta|$,
with subsequent damping and hardening as the system returns to its equilibrium state.
These amplitude, or ``Higgs'', oscillations were observed using tr-ARPES in 
charge- and spin-density wave systems\cite{schmitt_08,perfetti_08,ruegg_08,yusupov_10},
as well as cold atomic gases\cite{endres_12}.
\sout{The concept of perturbing the free energy landscape is regularly applied in cold atomic
systems, where the time scales are much slower, and have led to the identification
of the Higgs mode there\cite{endres_12}.}

%

\begin{figure*}[htpb]
	\includegraphics[width=0.8\textwidth]{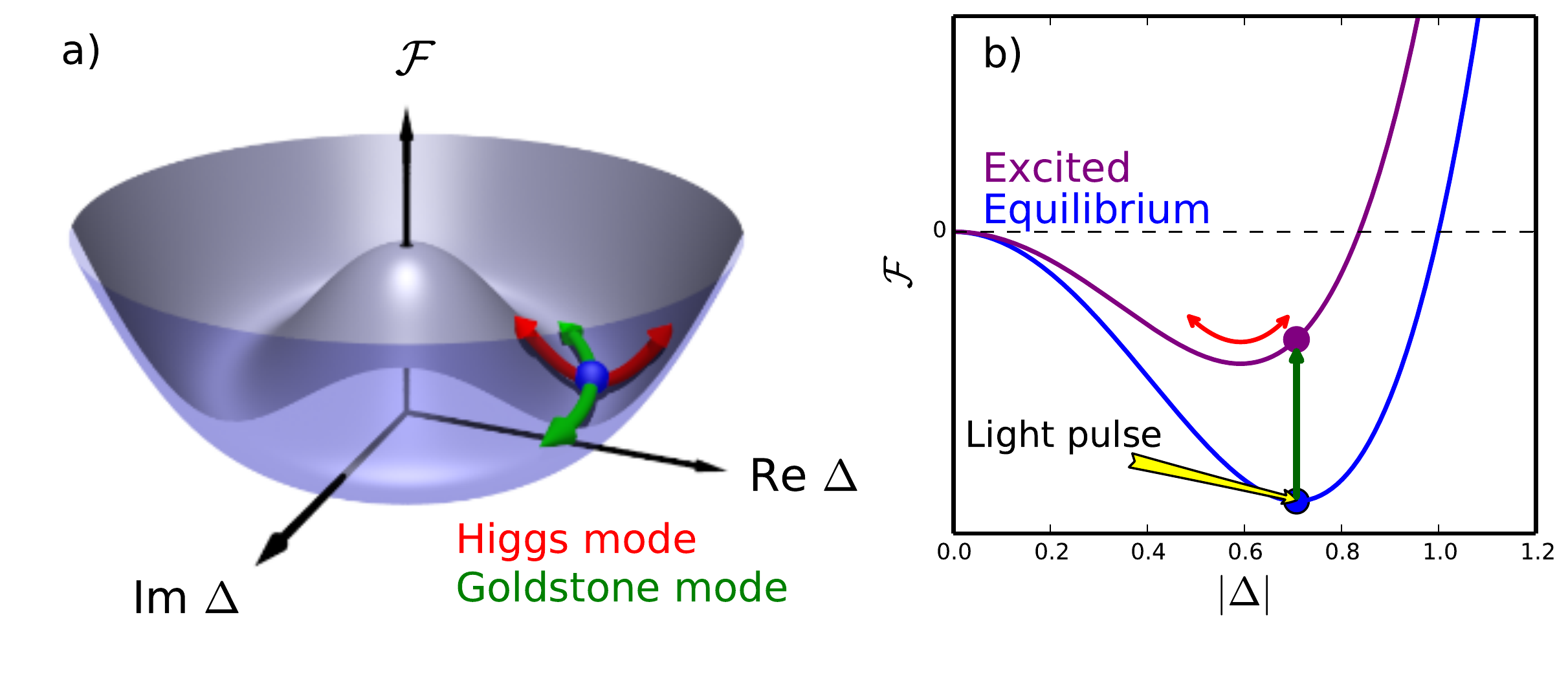}
	\caption{\textbf{Illustration of Higgs excitation} a) Free energy landscape $F[\Delta]$ for the complex order parameter near the phase transition. The Higgs (amplitude) and Nambu-Goldstone (phase) modes are indicated. b) Schematic of the excitation process. The light pulse displaces F on sufficiently short time scales such that the order parameter $|\Delta|$ cannot respond adiabatically leading to oscillations about the new minimum.}
	\label{fig:cartoon}
\end{figure*}

The dynamics of the superconducting
order parameter are often studied with
a ``quantum quench'', where one of the physical parameters that make up the
superconducting state is changed\cite{volkov_74,yuzbashyan_06},
resulting in damped oscillations with constant final frequency $2\Delta_\infty$.
A similar result was obtained numerically
through truncated equation of motion approaches.\cite{papenkort_07,papenkort_08}
Here, we go beyond the single-time approaches and
directly simulate the pump/probe process in a superconductor by self consistently
solving the Gor'kov equations in the time domain.
We simulate the pump-probe process
using a two-time Green's function formalism.
This formalism captures the full frequency dependence of the pairing interactions
the return to equilibrium through electron-phonon scattering,
and allows for
the calculation of the time-resolved single-particle spectral function, as measured with
tr-ARPES\cite{sentef_13}.
We focus on the spectra near the Fermi level, where the signatures of the order parameter $\Delta$
clearly appear both in and out of equilibrium.
After pumping, the system exhibits oscillations at twice the gap energy, which
is now time-dependent ($\Delta(t))$.  
As the pump fluence is increased, the gap
partially melts,
leading to slower oscillations.

An intuitive understanding of the dynamics that occur during the pump-probe process can be gained by
considering the free energy landscape for the complex ($U(1)$) order.  The ordered state is perturbed from its equilibrium
position by a laser pulse, resulting in changes of the free energy landscape $\mathcal{F}$.  The minimum in $\mathcal{F}$ is
reduced due to the decrease of quasiparticles involved in ordering.  If the response of the complex order parameter
is slower than the changes in $\mathcal{F}$, oscillations about the new minimum will occur (see Fig.~\ref{fig:cartoon})
at a frequency of $2|\Delta|$, with subsequent damping and hardening as the system returns to its equilibrium state.

\section{Methods and model}

We consider the Holstein model
\begin{align}
\mathcal H= \sum_\kk \epsilon(\kk) c^\dagger_\kk c_\kk + \Omega  \sum_i b_i^\dagger b_i - 
		g \sum_i c_i^\dagger c_i \left( b_i + b_i^\dagger \right)  \nonumber
\end{align}
where the individual terms represent the kinetic energy of electrons with a dispersion $\epsilon(\kk)$, the energy of phonons with a frequency $\Omega$, and a coupling between them whose strength is given by $g$.  Here, $c^\dagger_\alpha (c_\alpha)$ are the standard creation (annihilation) operators for an electron in state $\alpha$; similarly, $b^\dagger_\alpha (b_\alpha)$ creates (annihilates) a phonon in state $\alpha$.  For concreteness we study a square lattice dispersion with nearest neighbor hopping ($V_{nn})$,
\begin{align}
\epsilon(\kk) = &-2 V_{nn} \left[ \cos(k_x) + \cos(k_y)\right] -\mu
\end{align}
where $\mu$ is the chemical potential.  We have used the convention that $\hbar=c=e=1$, which makes inverse energy the unit of time.  

The electron-phonon interaction is treated at the self-consistent Born level,
where the self-energy is given by
\begin{align}
\bar\Sigma^\mathcal{C}(t,t') = i g^2 \bar\tau_3\ \bar G_\mathrm{loc}^\mathcal{C}(t,t') \bar\tau_3\ D^\mathcal{C}_0(t,t'),
\end{align}
where $\bar\tau_3$ is the $z$ Pauli matrix in Nambu space,
and $\bar G^\mathcal{C}_\mathrm{loc}(t,t') = \sum_\kk \bar G^\mathcal{C}_\kk(t,t')$ 
i.e. the local Green's function.

The equations to be solved are computationally demanding, and as such the parameters
are chosen with an eye towards the feasibility of the simulation.  We study a
square lattice tight-binding model at half filling.
The nearest neighbor hopping strength is $V_{nn} = 0.25$ eV, 
and the phonon frequency and coupling are chosen as $0.8V_{nn}$ and $1.38 V_{nn}$, respectively.
The resulting phonon coupling is of intermediate strength ($\lambda\approx 0.58$), which is
within the Migdal limit.
In the calculations, in addition to the strongly coupled Einstein phonon, to avoid unphysical metastable
states within the phonon window due to infinitely long lifetimes we include a weakly coupled low-energy phonon
in the distribution with $\Omega_\mathrm{weak}=1$ meV and $g^2_\mathrm{weak}=1$ meV$^2$.
For these parameter, the transition temperature $T_c \approx 18.7$ meV.  All data shown are calculated at $T=0.4 T_c$.

The pumped superconductor is modeled by self-consistently
solving the Gor'kov equations for the Migdal-Eliashberg model
in the time domain (see the SI for a detailed discussion).  
By treating superconductivity at this level, one avoids the issues of gauge invariance that arise in
isotropic attractive-U models.\cite{AGD, Rammer}
The electron-phonon interactions are treated on the level of the self-consistent Born approximation.
The resulting time-domain Green's functions are then used to obtain
the tr-ARPES spectra\cite{freericks_09}.

The solution of the Nambu-Gor'kov equations in the time domain requires self-consistency
on the entire Keldysh contour.
We utilize the standard two-time Keldysh formalism, where the 
contour Green's functions are 2x2 matrices in Nambu space\cite{AGD,Rammer},
\begin{align}
\bar{G}^\mathcal{C}_\kk(t,t') &= -i\left\langle\mathcal{T_C}
\left(
\begin{array}{cc}
c_{\kk\uparrow}(t) c^\dagger_{\kk\uparrow}(t') & c_{\kk\uparrow}(t) c_{-\kk\downarrow}(t') \\
c^\dagger_{-\kk\downarrow}(t) c^\dagger_{\kk\uparrow}(t')  & c^\dagger_{-\kk\downarrow}(t)  c_{-\kk\downarrow}(t')
\end{array}
\right)\right\rangle \\
&\equiv
\left(
\begin{array}{cc}
G^\mathcal{C}_\kk(t,t') & F^\mathcal{C}_\kk(t,t') \\ 
F^{\dagger\mathcal{C}}_\kk(t,t')  & -G^\mathcal{C}_{-\kk}(t',t) 
\end{array}
\right),
\end{align}
where $t$ and $t'$ lie on the Keldysh contour, and $\mathcal{T_C}$ is the contour time-ordering operator.
The equations of motion (on the contour) are
\begin{align}
\left(i\partial_t \bar\tau_0 - \bar\epsilon_\kk(t) \right) \bar{G}^\mathcal{C}_\kk(t,t') 
&= \delta^\mathcal{C}(t,t') \bar\tau_0 \nonumber \\
&+ \int_\mathcal{C}dz\  \bar\Sigma^\mathcal{C}(t,z) \bar G_\kk^\mathcal{C}(z,t')
\end{align}
\begin{align}
\bar\epsilon_\kk(t) = \left(
\begin{array}{cc}
\epsilon_\uparrow(\kk-\A(t)) & 0 \\
0 & -\epsilon_\downarrow(-\kk-\A(t))
\end{array}
\right)
\end{align}
where 
$\bar\tau_0$ is the identity matrix, $\epsilon_\uparrow(\kk) = \epsilon_\downarrow(\kk) = \epsilon(\kk)$ is the
bare dispersion given above, and $\A(t)$ is the time-varying vector potential in the Hamiltonian gauge.
On the Keldysh contour, the Langreth rules can be applied to separate the contour equation into various
well-known components: the Matsubara ($M$), lesser ($<$), and greater ($>$) Green's functions, 
as well as the mixed real-imaginary $\rceil/\lceil$ types.
The equations of motion are solved on the contour by using massively parallel computational
methods for integro-differential equations, as described in 
Ref.~\onlinecite{stefanucci_book}.
The data in the normal state is obtained by performing the simulations without allowing a solution in the anomalous
channel.

Once the Green's functions are obtained, time-resolved ARPES (tr-ARPES) spectra can be computed.  For a probe pulse of width $\sigma_p$, the gauge-invariant tr-ARPES intensity at time $t_0$ is \cite{freericks_09}
\begin{align}
I(\kk,\omega,t_0) =\mathrm{Im} \int dt\, dt'\, 
p(t,t',t_0)
e^{i\omega(t-t')} G_{\tilde\kk(t,t')}^<(t,t') 
\label{eq:trarpes}
\end{align}
where $p(t,t',t_0)$ is a two-dimensional normalized Gaussian with a width $\sigma_p$ centered at $(t,t')=(t_0,t_0)$.  Note that here, only one component of the full Nambu matrix is used.
The field-induced shift in $\kk$  has to be corrected via a gauge shift in the momentum argument of $G^<_\kk$ with \cite{v_turkowski_book}
\begin{align}
\tilde\kk(t,t') = \kk + \frac{1}{t-t'} \int_{t'}^{t} d\bar t\, \A(\bar t).
\end{align}
To determine the tr-ARPES spectral weight, we utilize a probe with a Gaussian envelope whose width $\sigma_p=16.45$ fs.

\begin{figure*}[htpb]
	\includegraphics[clip=true, trim=0 0 0 5,width=0.99\textwidth]{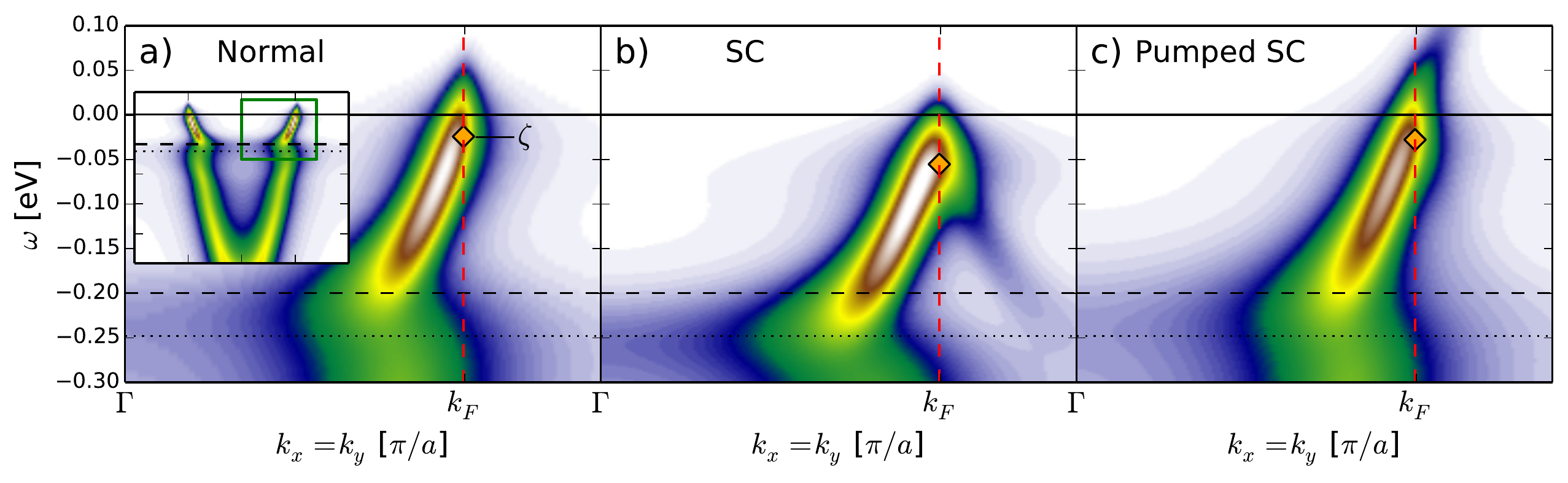}
	\caption{tr-ARPES spectra of a) the normal state in equilibrium 
	(the inset shows the full spectrum, with the green
	box indicating the region for panels a-c),
	b) the superconducting state in equilibrium, and
	c) the superconducting state, $65.8$ fs after a pump with $E_\mathrm{max} = 0.9$ V/$a_0$.
	The vertical red lines
	indicate the Fermi momentum $k_F$ ,
	the horizontal solid (dashed) line indicates the Fermi level (phonon frequency $\Omega$).
	The dotted horizontal line indicates the gap-shifted phonon frequency ($\Omega+\Delta_0$).
	The maximum along the line $\kk=k_F$, denoted $\zeta$, is shown with a marker.
	}
	\label{fig:fig1}
\end{figure*}

The field is explicitly included via the Peierls'
substitution $\kk(t) = \kk-\A(t)$, where $\A(t)$ is the vector potential in the Hamiltonian gauge,
which has no scalar potential $\Phi$.
This includes the field beyond linear coupling, which is critical for the excitation
of the Higgs mode (see Ref.~\onlinecite{matsunaga_14}).

We have checked that the inclusion of local electron-electron scattering up to second order in the interactions does not qualitatively affect our results (a comparison is shown in Sec.~\ref{sec:e-e}),
where the addition to the self-energy is
given by
\begin{align}
\bar\Sigma^\mathcal{C}(t,t') = U^2 \bar\tau_3\ \bar G_\mathrm{loc}^\mathcal{C}(t,t') \bar\tau_3\ 
\mathrm{Tr}\left\{ \bar G_\mathrm{loc}^\mathcal{C}(t,t') \bar\tau_3 \bar G_\mathrm{loc}^\mathcal{C}(t',t) \bar\tau_3 \right\}.
\end{align}

The phonon bath is kept fixed by ignoring the feedback of the electrons on the phonons;
we remain outside of the regime where this is expected to be important, i.e.
strong pumping, the formation of static Peierls distortions (CDWs), 
or quenches of the interaction constant\cite{murakami_14}.

\section{Results}
Figure~\ref{fig:fig1} shows the tr-ARPES spectra obtained from the calculations 
in equilibrium (a, b) and after pumping (c).  
Here, the maximum field strength $E_\mathrm{max} = 0.9$ in
volts per lattice constant (V/$a_0$).
The spectra are broader than is usual in equilibrium ARPES due to the Heisenberg uncertainty introduced
by a time-resolved measurement.
The diamond markers indicate the maximum of the constant momentum cut (energy distribution curve
or EDC) at $\kk=k_F$, which we will denote by $\zeta$ throughout.
In equilibrium, the spectra show the usual hallmarks of a strongly coupled BCS superconductor:
the spectra are pulled back from the Fermi level by some amount, visible both in the decrease of spectral
weight at the Fermi level, as well as the downward shift of $\zeta$.
In addition, the kink due
to the strongly coupled phonon at $\omega=\Omega$ shifts down in binding energy, and 
shadow bands appear along $\omega=-\epsilon_\kk$ due to the particle-hole mixing.
For the equilibrium superconductor $\zeta$ is at roughly $-55$ meV, although the magnitude
of the true gap $\Delta_0$, 
which is determined from the spectral gap in the 
equilibrium self-energy,
is slightly less ($\Delta_0\sim 48$ meV).
The difference arises because $\zeta$ is shifted by the
broadening of the single-particle spectrum and probe resolution.

\begin{figure*}[htpb]
	\includegraphics[width=\textwidth]{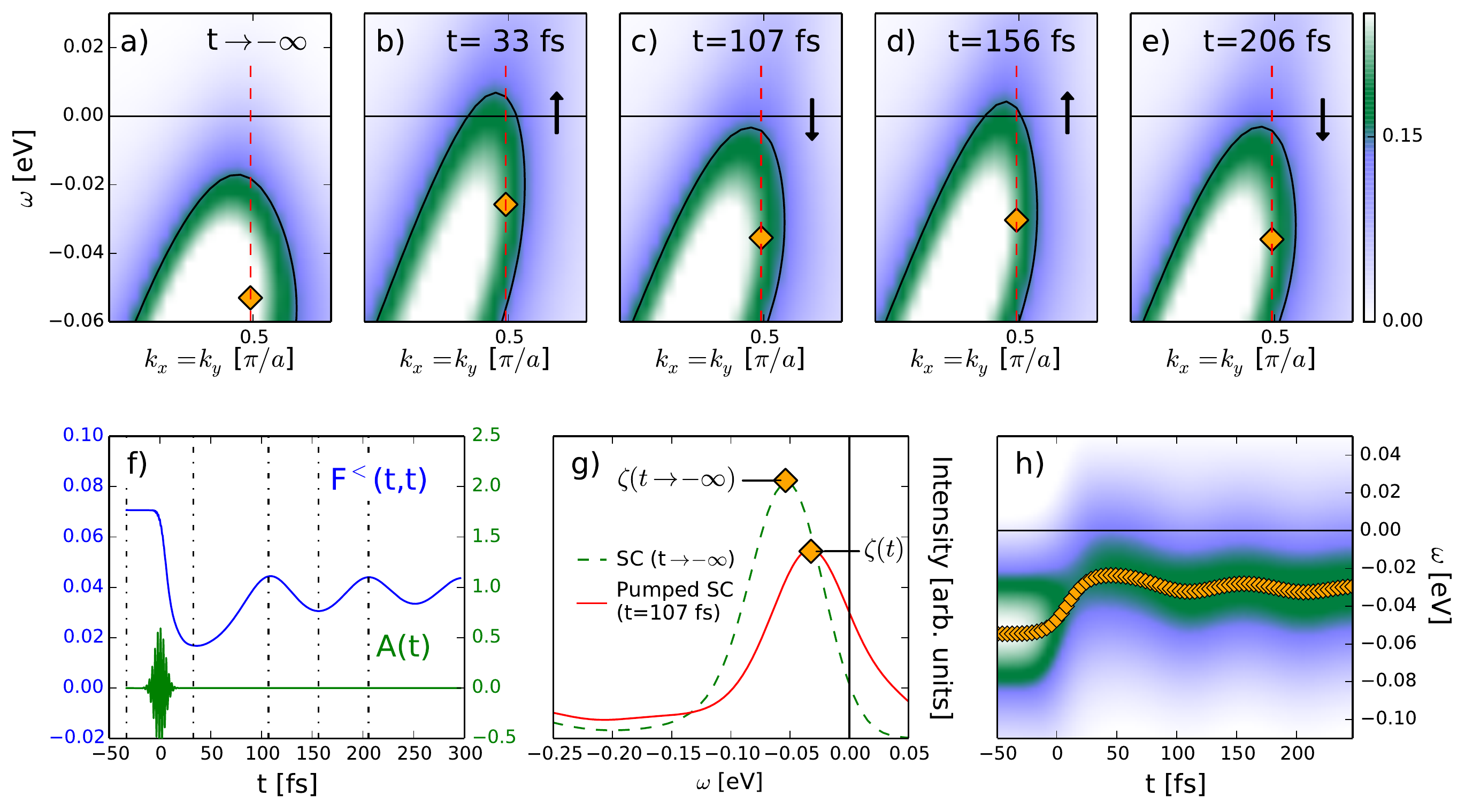}
	\caption{
	a) tr-ARPES spectum near the Fermi level in equilibrium. 
	b)-e) Spectra at various times after the pump, illustrating the shift of the spectral weight
	back and forth across the Fermi level
	(a movie is available showing this in detail in the supplementary information).
	The maximum along the (red) line $\kk=k_F$, denoted $\zeta$, is shown with a marker.
	f) Anomalous density $F^<(t,t)$ (see text),
	which measures the strength of the superconducting state, and the
	vector potential [$A(t)$] as a function of time.  
	The vertical lines indicate the time slices shown in panels
	a-e. The black arrows indicate the direction of the shift of the spectrum from the previous panel.
	g) Energy distribution curves (EDCs) at $k=k_F$ (red lines in panels a-e) 
	for equilibrium
	and pumped superconductor ($t=107$ fs).  The marker indicates the EDC maximum (at $\omega=\zeta$). 
	h) False-color plot of EDC intensity as a function of time showing the oscillations after the pump.
	}
	\label{fig:fig2}
\end{figure*}

After pumping, the characteristics of superconductivity are reduced in magnitude.  $\zeta$ and
the gap-shift of the phonon kink 
are reduced, and the spectral weight in the shadow bands is no longer
visible.  In fact, the spectrum more closely resembles a normal metal at some elevated temperature.
However, superconductivity never fully disappears at this field strength (as we will show below),
suggesting that an elevated temperature scenario does not fully capture the behavior here.

We next utilize the strength of the tr-APRES approach and
focus on the changes of the spectra 
near the Fermi level ($E_F$) where
the signature of SC is strongest
(Figure~\ref{fig:fig2}).
The figure shows snapshots of the tr-ARPES spectra before (\ref{fig:fig2}a), and after
(\ref{fig:fig2}b-\ref{fig:fig2}e) pumping.

Immediately after the pump (Fig.~\ref{fig:fig2}b), there is a shift of spectral weight from
lower binding energies to
the Fermi level,
partially closing the gap,
and the EDC center $\zeta(t)\ (\Diamond)$ shifts back towards the Fermi level.
The following panels show
that the spectral gap first closes right after the pump (\ref{fig:fig2}b) and then reopens and closes
in the successive time slices (\ref{fig:fig2}c-\ref{fig:fig2}e), leaving behind a gap which is slightly reduced
compared to its equilibrium value (\ref{fig:fig2}e). Thus both the spectral intensity as well as the spectral
maximum oscillate in time after the pump is applied.

\begin{figure}[htpb]
	\includegraphics[clip=true, trim=0 0 0 -10,width=0.99\columnwidth]{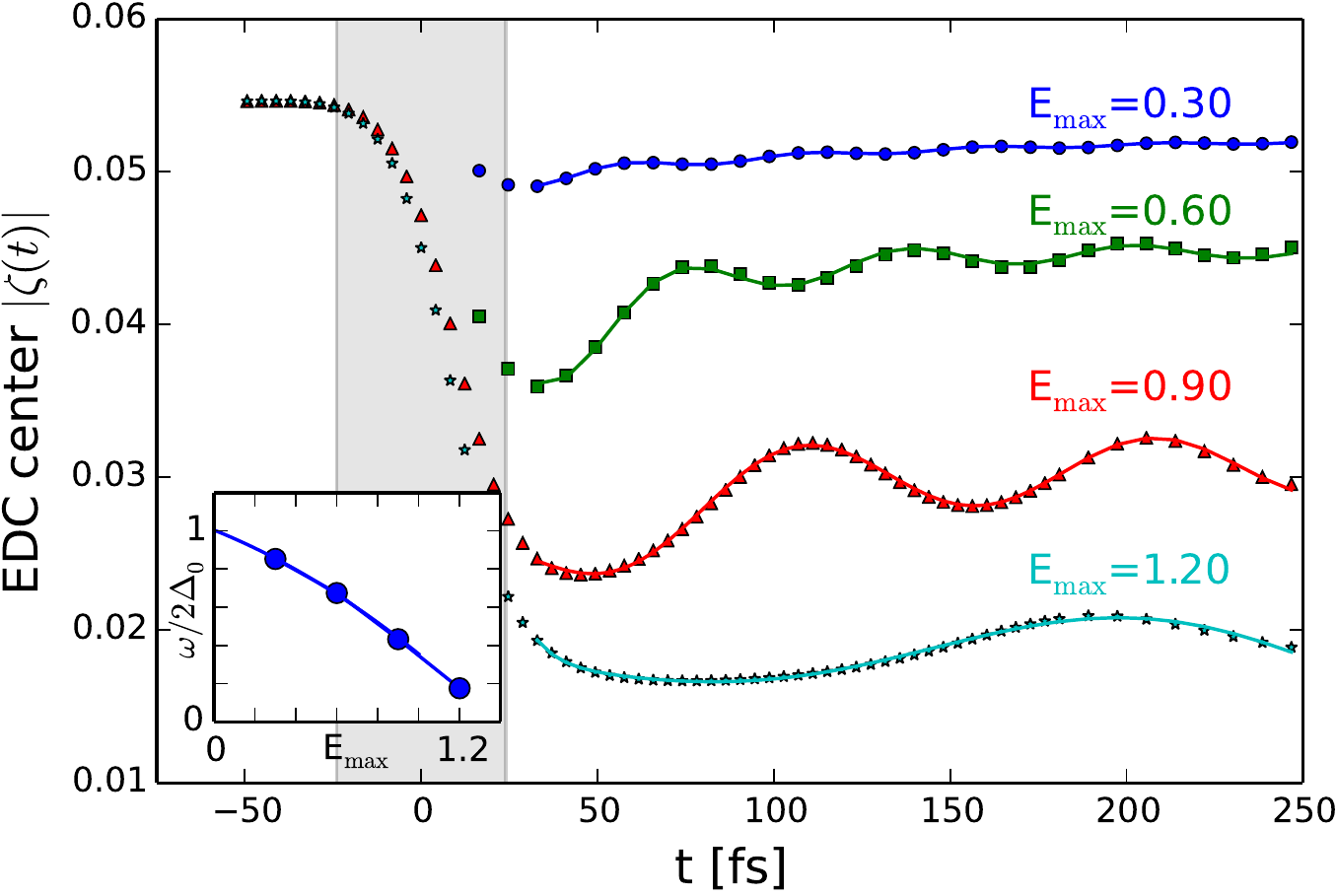}
	\caption{
	The position of the EDC maxima ($\zeta(t)$)for various pump
	fluences.  The shaded region indicates the times for which the pump field is on (as defined
	by 1.5 the field width $\sigma$).  Solid lines are
	fits to a decaying exponential plus a damped oscillation.
	Inset: fitted oscillation frequencies as a function of pump fluence
	(maximum field in V/$a_0$).  The solid line is a quadratic polynomial fit.
	}
	\label{fig:zeta_vs_t}
\end{figure}

To show that the superconductivity remains even though there is spectral weight in the gap,
we consider the ``anomalous density'' $F^<(t,t) \equiv \sum_{\kk} F_\kk^<(t,t)$
[in analogy with the normal density $n(t) \equiv -i \sum_\kk G_\kk^<(t,t)$]  
shown in panel f. 
In equilibrium BCS theory, this quantity is related to one side of the gap equation,
\begin{align}
F^<_\kk(t,t'=t) = \frac{\Delta_0}{2E_\kk} \tanh\left( \frac{E_\kk}{2T} \right),
\end{align}
where $E_\kk = \sqrt{\epsilon_\kk + \Delta_0^2}$.
After pumping, although the magnitude of the order parameter is reduced,
superconductivity is still present.  Moreover, $F^<(t,t)$ shows the same oscillations as observed
in the spectra.  
The oscillations occur for long times after the pump
pulse
indicating that they are intrinsic to the superconducting state, rather than
directly related
to particulars of the pump.
The snapshots (panels c-e) are taken at times corresponding
to the minima and maxima of the oscillations.

\begin{figure}[htpb]
	\includegraphics[clip=true, trim=10 10 0 5,width=0.99\columnwidth]{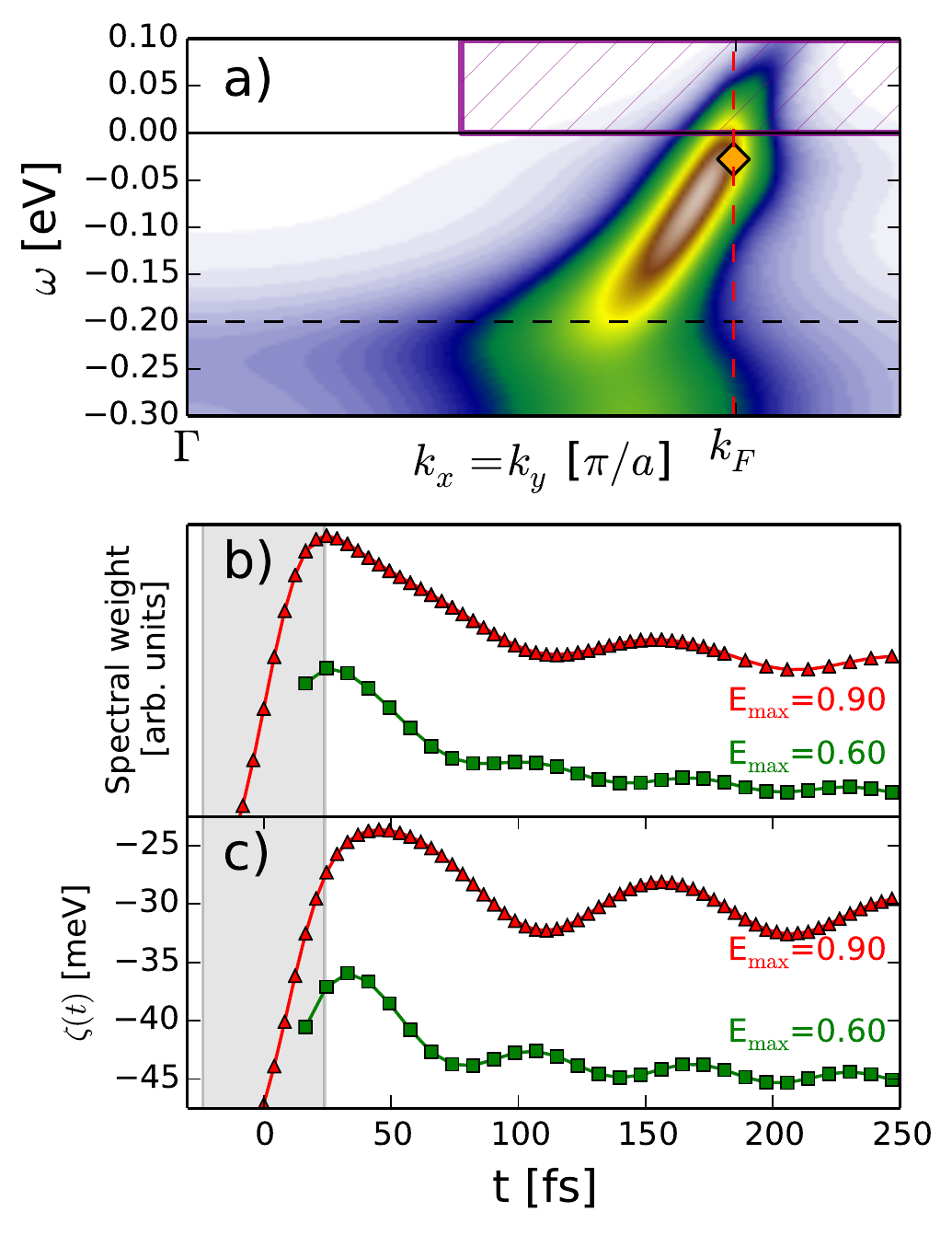}
	\caption{
	a) tr-ARPES spectrum in the superconducting state, $65.8$ fs after the pump with
	E$_\mathrm{max} = 0.6$ V/$a_0$.
	The hatched area indicates the region for integration in panel b), the orange marker the EDC maximum.
	b) Spectral weight integrated above $E_F$ for two pump fluences (indicated in V/$a_0$).
	c) EDC maxima $\zeta(t)$ reproduced from Fig.~\ref{fig:zeta_vs_t}.
	}
	\label{fig:sw_above_ef}
\end{figure}

We further investigate the oscillations by considering the EDCs
at $\kk=k_F$ and analyze the dynamics.
Figure~\ref{fig:fig2}g) shows the EDCs for the equilibrium and pumped superconductor
(at $t=107$ fs).  
After pumping, $\zeta$ returns towards the Fermi level, but not fully.
Figure~\ref{fig:fig2}h) shows the EDCs as a function of
time delay.
Upon arrival of the pump, the superconductivity is reduced as spectral
weight is scattered to above the Fermi level and across the Brillouin zone.  
The band subsequently shifts back and forth at a particular frequency.
This is markedly different from the normal state,
where the spectra after pumping 
return monotonically to equilibrium\cite{sentef_13} (unless phonons are resonantly excited or
the pump is sufficiently strong\cite{murakami_14}),
indicating that the superconducting order is responsible for the oscillations.

By increasing the pump fluence, the order can be further reduced, and the effects of further reduction
on the oscillations can be observed.
Fig.~\ref{fig:zeta_vs_t} shows the oscillations after pumping for 
various pump fluences.
With increasing pump fluence, the SC is further suppressed, as reflected in the reduction of $|\zeta|$.
Concomitantly with the decrease in SC, the oscillations slow down,
confirming that the mode is determined by the state of the system after pumping.
This is the same mechanism that leads to changes in the effective self-energy in the
normal state after pumping\cite{kemper_14}, although the field threshold where the system
deviates from the equilibrium behavior is much lower.
$\zeta(t)$ is fitted to a decaying exponential combined with a damped oscillation.  
The obtained frequencies $\omega$, scaled by twice the equilibrium gap $2\Delta_0$, are
shown in the Figure inset. 
In the limit of zero fluence, the oscillation frequency $\omega$ extrapolates to $2\Delta_0$, 
the equilibrium gap.
This implies that
tr-ARPES can provide a clean measurement of $\Delta_0$, which is
obscured in equilibrium by broadening of the spectral function and energy resolution.

\begin{figure}[htpb]
	\includegraphics[width=\columnwidth]{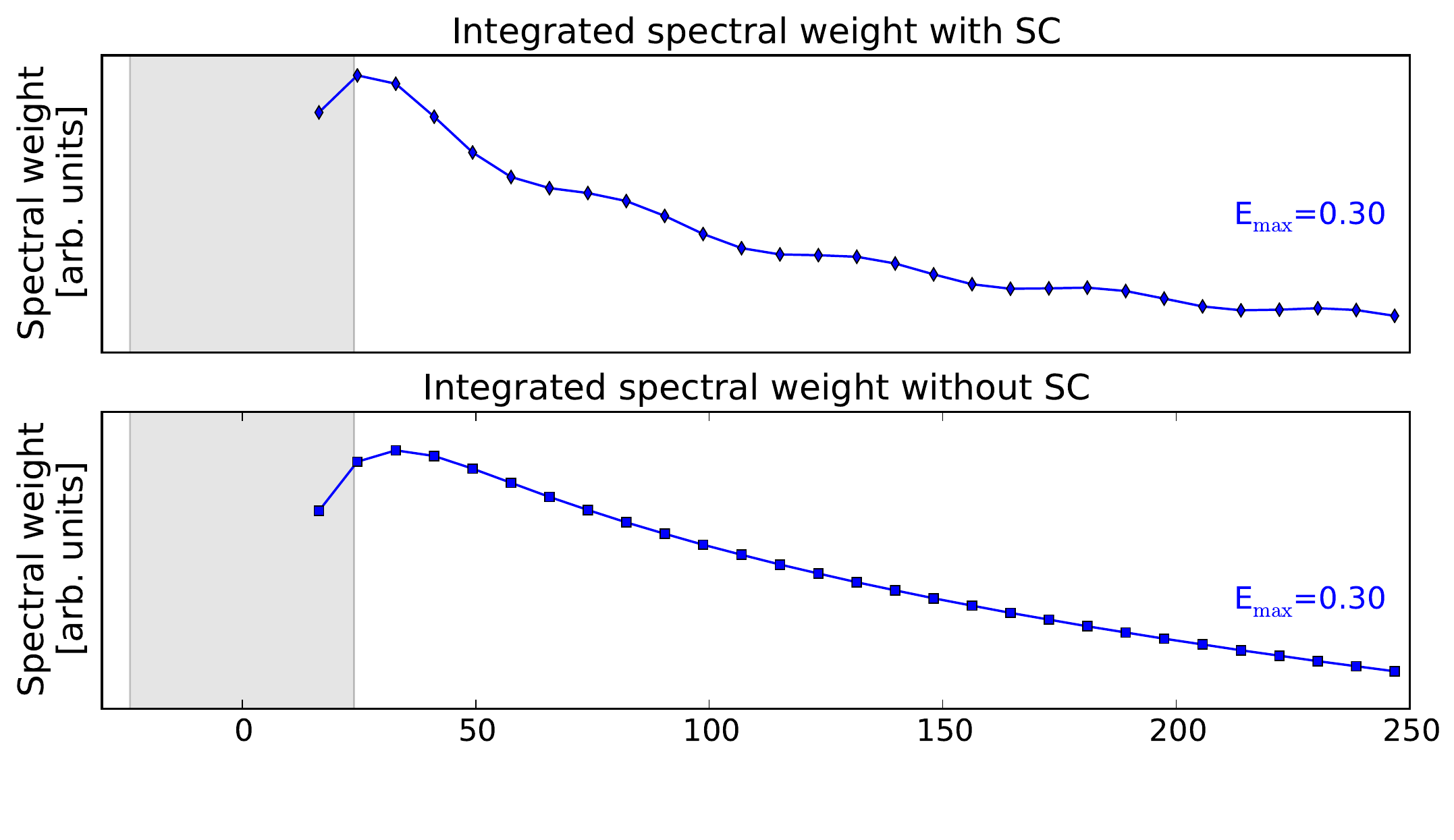}
	\caption{Spectral weight integrated above $E_F$ for comparing the superconducting and normal states
	(indicated in V/$a_0$). The gray region indicates the times where the field is on.  The oscillations are only visible
	when the superconducting order is finite.}
	\label{fig:sw_above_ef_nosc}
\end{figure}

The effects of the changing gap at the Fermi level
due to amplitude mode oscillations are visible across the
entire spectrum, including at the gap edge, at the phonon kink, and above the Fermi level.
To illustrate this, we integrate the spectral weight above the Fermi level, where the 
experimental signal to noise ratio is usually large.
Fig.~\ref{fig:sw_above_ef} compares the integrated spectral weight (Fig.~\ref{fig:sw_above_ef}b) with $\zeta(t)$ extracted
from the EDCs as before.  The oscillations are readily resolvable in both cases, in particular for weaker fields
where the early-time behavior is not dominated by simple scattering.  To further underscore the point that these oscillations
are absent when there is no superconducting condensate, Fig.~\ref{fig:sw_above_ef_nosc} shows a comparison for similar
fields between the superconducting and normal states.  

\section{Effect of electron-electron interactions}
\label{sec:e-e}

\begin{figure}[htpb]
	\includegraphics[width=\columnwidth]{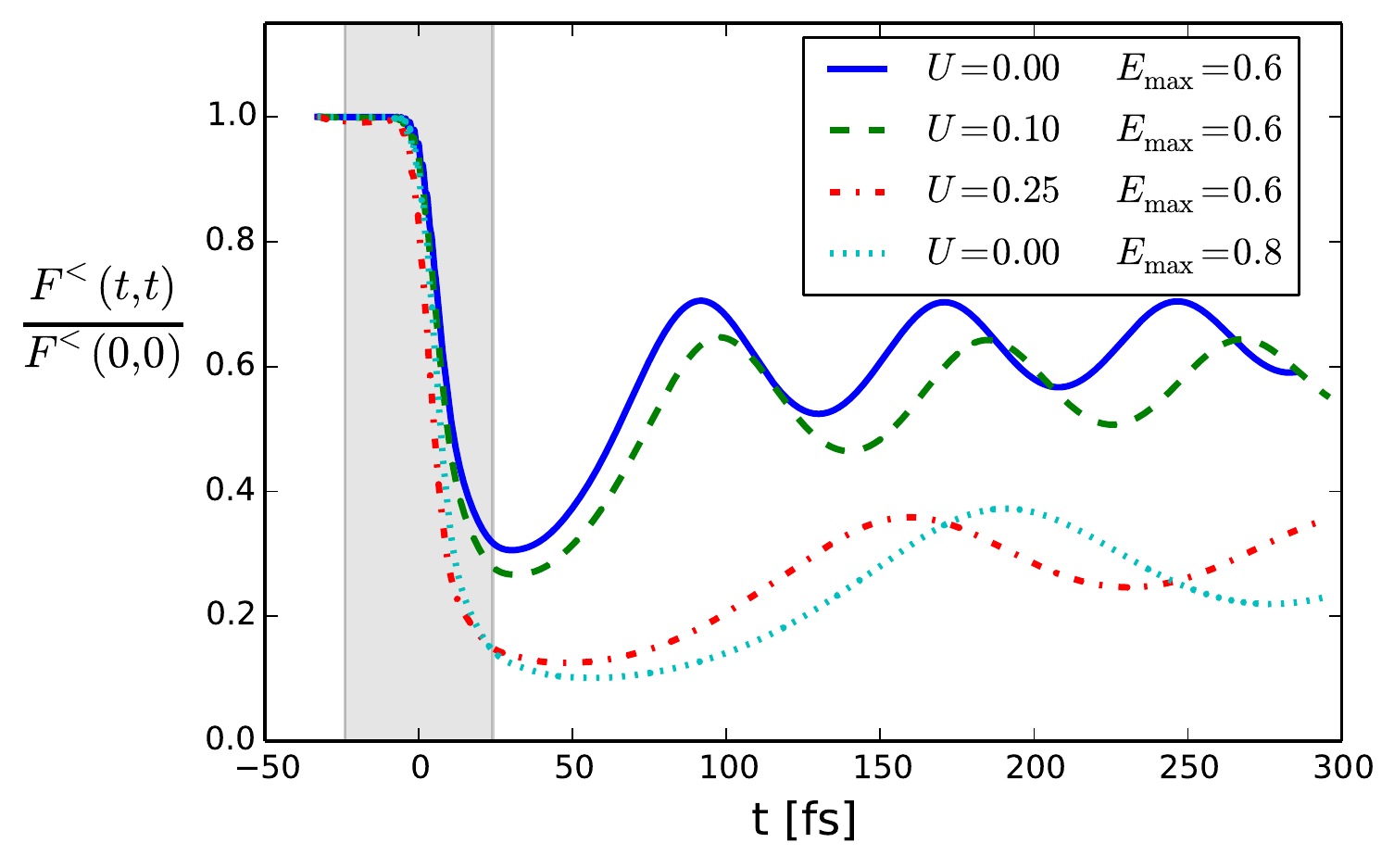}
	\caption{
	Normalized ``superconducting density'' $F^<(t,t)$ (see main text) for various strengths
	of the Coulomb scattering $U$.  The gray region indicates the times where the field is on.   Fields
	are in units of V$/a_0$, and interactions in eV.
	}
	\label{fig:ee}
\end{figure}

To account for electron-electron interactions, we have considered
local electron-electron
scattering ($U$) at the level of self-consistent second order perturbation theory. The inclusion of
electron-electron scattering changes the energy absorption and thus the state after pumping, leading to a
change in the oscillation frequency as the superconducting order is weaker, as illustrated in Fig.~\ref{fig:ee}.
For comparison, a data set with larger field but without Coulomb scattering is also shown.

\section{Conclusion}
The results of this study indicate that time-resolved ARPES can be used to directly study the dynamics
of Cooper pairs by examining the time-resolved
single particle spectral function,
making the identification and examination of the Higgs mode
available in superconductors.
This opens up new avenues for studying superconductivity, both BCS and unconventional. By perturbing
the superconducting order from its equilibrium state through nonlinear coupling to a strong field,
we can access regions of phase space
that are not sampled in equilibrium.  This could be particularly interesting in the case where several
competing orders are present, such as in the high-T$_c$ cuprates and pnictides.

More generally, the field of pump-probe spectroscopy, and nonlinear coupling to a driving potential,
is providing the means to observe
phenomena that were previously inaccessible to experiment. Here, we have illustrated this concept
in the context of Higgs oscillations in condensed matter, and observations were reported previously in 
two rather dissimilar systems: NbN\cite{matsunaga_13,matsunaga_14}, and cold atomic gases\cite{endres_12}.

The Higgs mode is just one example of a phenomenon that can
be observed or unraveled from others by going into the time domain and perturbing the
system far from its equilibrium state.
Others include, for example, pumping the lattice and driving the system from the disordered
to a (possibly) ordered phase\cite{mankowsky_14}.
Within this context, the
combined capability of experiment and theory in non-equilibrium spectroscopy and
modeling is set to grow into a fruitful approach to studying emergent physics.

\begin{acknowledgments}
We would like to thank P. Kirchmann for helpful discussions. 
A.F.K. was supported by the Laboratory Directed Research and Development Program of Lawrence Berkeley
National Laboratory under U.S. Department of Energy Contract No. DE-AC02-05CH11231.
B.M and T.P.D. were supported by the U.S. Department of Energy, Office of Basic Energy Sciences, Materials Sciences and Engineering under Contract No. DE-AC02-76SF00515.
J.K.F. was supported by the U.S. Department of Energy, Office of Basic Energy Sciences, Materials Sciences and Engineering under Contract
No. DE-FG02-08ER46542 and 
also by the McDevitt bequest at Georgetown.
The collaboration was supported by the U.S. Department of Energy, Office of Basic Energy Sciences, Materials Sciences and Engineering under Contract No. DE-SC0007091.
Computational resources were provided by the National Energy Research Scientific Computing Center supported by the U.S. Department of Energy, Office of Science, under Contract No. DE-AC02-05CH11231. 
\end{acknowledgments}
\bibliography{tdsc}

\end{document}